\documentclass[preprint,tightenlines,eqsecnum,floats,aps,amsmath,amssymb,nofootinbib,prd,showpacs]{revtex4}
\usepackage{mathtools}
\usepackage{multirow}
\usepackage{amssymb}
\usepackage{stmaryrd}
\usepackage{amsmath}
\usepackage{amsfonts}
\usepackage{mathrsfs}
\usepackage{CJK}
\usepackage{amsmath,amssymb,amsfonts}
\usepackage{graphicx}
\usepackage{appendix}
\def\be{\begin{equation}}
\def\ee{\end{equation}}
\def\ba{\begin{eqnarray}}
\def\ea{\end{eqnarray}}

\begin{document}

\title{Suitable resolution for EOS tables in neutron star investigation}

\author{Houyuan Chen} 
\affiliation{School of Physics and Optoelectronic Technology, South China University of
Technology, Guangzhou 510641, P.R. China}

\author{Dehua Wen\footnote{Corresponding author. wendehua@scut.edu.cn}} 
\affiliation{School of Physics and Optoelectronic Technology, South China University of
Technology, Guangzhou 510641, P.R. China}

\author{Na Zhang} 
\affiliation{School of Physics and Optoelectronic Technology, South China University of
Technology, Guangzhou 510641, P.R. China}

\date{\today}

\begin{abstract}
Inasmuch as the hydrostatic structure of interior neutron stars uniquely depends on the equation of state (EOS), the inverse constraints on EOS from astrophysical observation have been important methods to reveal the properties of the high-density matter. Up to date, most of EOSs for neutron-star matter are given in tabulations, but these numeric tables could be quite different in resolution. To guarantee both the accuracy and efficiency in computing Tolman-Oppenheimer-Volkoff (TOV) equation, a concise standard for generating EOS tables with suitable resolution is investigated in this work. It is shown that the EOS tables with 50 points logarithmic-uniformly located at supra-nuclear density segment [$\rho_{0}, 10 \rho_{0} $], where $\rho_{0}$ is the nuclear saturation density) would correspond to the interpolation-induced errors at $\sim 0.02\%$ for the gravitational mass $M$ and $\sim 0.2\%$ for the tidal deformability $\Lambda$.

\end{abstract}


\maketitle

\section{Introduction}

The pulsars in observation \cite{atnf,chandra,Manchester2005} are generally considered as neutron stars in theory. Most of the observed neutron stars have mass around 1.4 solar masses ($\rm{M}_{\odot}$) \cite{nsmasses}, and the PSR J0348+0432 is the heaviest observed neutron star with a precise mass measurement ($2.01 \pm 0.04 \, \rm{M}_{\odot}$) \cite{Antoniadis13}.
In theoretical perspectives, the macroscopic properties of neutron star, such as the maximum mass ($M_{max}$), depend strongly on the equation of state (EOS) of high-density matter \cite{Theoretic1,Theoretic2,Theoretic3}. The $M_{max}$ of neutron star delimits whether it is hydrostatic stable or it will finally collapse into a black hole through the oscillation process \cite{compactoj4}. The recent  constraints on the $M_{max}$ based on the GW170817 observation are concentrated around $2.20~\rm{M}_{\odot}$. For example, based on the GW170817 observation, three different groups have given the constraint on the maximum mass as $M_{max} < 2.17~\rm{M}_{\odot}$ \cite{TMmax}, $2.15~\rm{M}_{\odot} < M_{max} < 2.25~\rm{M}_{\odot}$ \cite{Shibata2017} and $2.16~\rm{M}_{\odot} < M_{max} < 2.28~\rm{M}_{\odot}$ \cite{Ruiz2018}, respectively.  According to the same gravitational waves (GW) observation, people gave the constraints on the radius of a canonical neutron star as $11.0 < R_{1.4}/ \rm{km} < 13.2$ \cite{RMrelation1,inves1}.

Lots of pioneering works have been done to predict the gravitational waves emitted from a binary neutron star (BNS) system \cite{GWT1,GWT2,Baiotti2017}. The numeric simulations of BNS mergers, which are various with different choices of the EOSs, have illustrated possibility for probing indirectly the properties of neutron-star matter from gravitational waves (GWs) \cite{GWS1,GWS2,GWS3}. Postnikov \emph{et al.} pointed out that the dimensionless tidal deformability $\Lambda$, which can be revealed from the GW signals during the coalescence process of BNS, is capable of characterizing different EOSs \cite{neutron starstructure3}. The first GW detection of BNS coalescence (GW170817)  put a   constraint on the tidal deformability as $\Lambda_{1.4}  < 800$ \cite{GW0} for canonical neutron stars from the first analysis. An improved analysis of the GW170817 provided both the upper and lower limits for the tidal deformability as ${\Lambda _{1.4}} = 190_{ - 120}^{ + 390}$, which leads to a constraint on the EOS at twice nuclear saturation density as $p(2{\rho _0}) = 21.85_{ - 10.61}^{ + 16.85} \, \rm{MeV/fm^3}$ \cite{GW1}. With observational constraints on $\Lambda$, Most \emph{et al.} generated millions of EOSs from their parameterized sets and then exploited more than $10^{9}$ equilibrium models for neutron stars to measure the typical radius ${R_{1.4}} = 12.39_{ + 1.06}^{ - 0.39}$ $\textrm{km}$ at $2\sigma $ level \cite{GW2}.

The theoretical determination of $\Lambda  = (2/3){({c^2}/G)^5}{(R/M)^5}{k_2}$ requires precise inner solutions of the Tolman-Oppenheimer-Volkoff (TOV) equation \cite{GWA,Tolman1939,Oppenheimer1939}. The relevant tidal Love number $k_2$ is determined by the hydrostatic distribution of the stars \cite{GWA}. The tidal deformability $\Lambda$ deduced from $k_2$ can be used to discriminate the EOSs.
On the other hand, there is no unified model yet to describe the EOS of the compressed matters \cite{Li2008,Lattimer2016,Feng2018}. Even in a specific model, it generally takes complex computation to provide the $\rho-\varepsilon-p$ relation, where $\rho$ is the baryon number density, $\varepsilon$ is the energy density and $p$ is the pressure. Therefore, numeric EOS tables become a convenient choice in neutron star study. For the realistic EOSs (that is, the tabular EOSs), the solutions of stellar structures have to be given by numerical integration.

In the integration process, the single-step errors must be restrained to provide accurate results of $k_2$. By contrast, different from the integration errors that can be handled simply with shorter step-sizes, the interpolation-induced errors are mainly affected by the resolution of EOS tables. Nonetheless, when employing a huge number of EOSs to investigate the neutron-star characteristics by Bayesian methods \cite{GW1,GW2,Annala2018,Greif2018}, the efficiency of the interpolation is of crucial important. As too large-size EOS tables are unpractical in the statistical study, a suitable resolution for EOS tables becomes particularly important.

This paper is organized as follows. In Sec.$\,$II, a brief introduction of  pragmatic techniques to deal with both the integration and the interpolation is given first. Then two widely-used interpolation methods are introduced to inspect the interpolation-induced errors from EOS tables. The minimal size of an EOS table to provide accurate results of $M$ and $\Lambda$ is also discussed in this section. In Sec.$\,$III, we extend the investigation onto different meshing methods for EOS tables to further examine the model-dependence of the discussion. At the end, a concise summary will be given.

\section{Numeric setup and EOS grid resolution}

 In general relativity, the structure of a static, non-rotating and spherical compact star is normally described by the TOV equations. The TOV equations can be written as \cite{Tolman1939,Oppenheimer1939},
\begin{equation} \label{TOV1}
\frac{{dp}}{{dr}} =  - \frac{{G(\varepsilon + \frac{{p}}{{{c^2}}})(m(r) + \frac{{4\pi {r^3}p}}{{{c^2}}})}}{{{r^2}(1 - \frac{{2Gm(r)}}{{r{c^2}}})}},
 \end{equation}
 \begin{equation} \label{TOV2}
  \frac{{dm(r)}}{{dr}} = 4\pi \varepsilon {r^2},
\end{equation}
where  $m(r)$ refers to the gravitational mass within  radius $r$,  and $G$ and $c$ are the gravitational constant and the light speed, respectively.

An important method to investigate the macroscopic properties of neutron stars is to numerically integrate the Eqs. (\ref{TOV1}) and (\ref{TOV2}) from the center ($m=0$, $r=0$, $\varepsilon = \varepsilon_c$) to the surface ($p=0$, $r=R$ and $m(R)=M$). The widely used forth-order Runge-Kutta (RK-4) method \cite{DefinitionRK0,DefinitionRK1} is applied as the high precision integration algorithm in this work. Moreover, we denote the relative deviation of a quantity $Q$ as $resQ$ ($=\left| Q-Q_T \right|/Q_T$, where $Q_T$ is the exact value) to discuss the precision issue.

\begin{figure}
\centering
\includegraphics [width=0.6\textwidth]{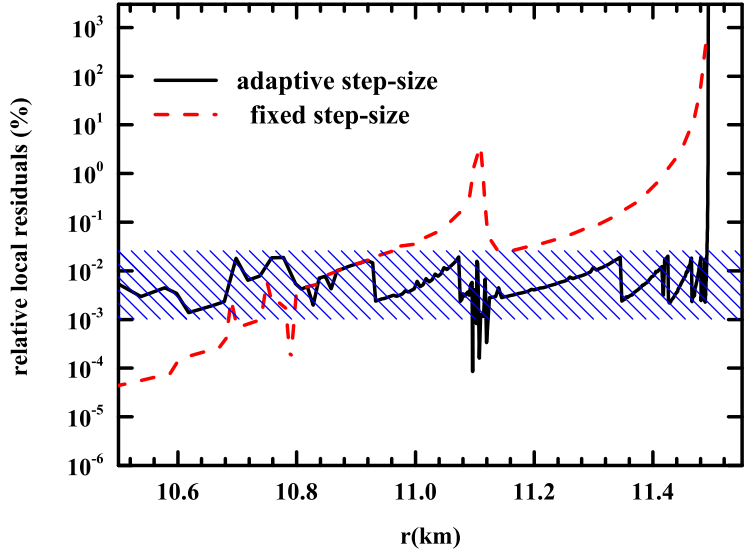}
\caption{ \label{resP} The percentage of local residuals for fixed step-size and adaptive step-size. The percentage of local residuals are defined as the errors of the pressure increments in each step divided by the precise values of pressure increments. The dash line represents the fixed step-size ($h=10\textrm{(m)}$) while the solid line denotes the adaptive step-size.  Here the APR EOS is employed and the central density  is  $4 \rho_0$.}
\end{figure}

The RK-4 method with adaptive step-size control \cite{emb1,emb2,emb3} can guarantee the global accuracy of the radius. In a fixed step-size computation, the local errors at the outer layers can increase rapidly when integrating outwards, which have been shown in Fig.$\,$\ref{resP}. The residuals for the final outcomes of radius are strongly relevant to these errors. The adaptive method effectively controls the local integration errors at outer crust layer until several meters to stellar surface. This technique is expected to improve computational precision significantly at crust which is of great concern in the investigations of low-mass stars.  With a proper method to integrate, the induced radial errors should be handled to $< 0.1\%$.

Apparently, the most important input in solving Eqs. (\ref{TOV1}) and (\ref{TOV2}) is the $\varepsilon - p$ relation. Currently, EOSs in tabulation form are the most common way in neutron-star investigations. To properly use EOS tables, it is necessary to apply interpolation method to obtain the intermediate values. The errors generated from the interpolation process could be quite different according to both the specific method of interpolation and the resolution of EOS table.

A simple but non-rigorous approximation for EOS table could be piecewise polytropic,
\begin{equation} \label{polytropicEOS}
{\log _{10}}p = {\log _{10}}K + \gamma \left( {{{\log }_{10}}\varepsilon  - {{\log }_{10}}{\varepsilon _0}}  \right),
\end{equation}
where $\varepsilon_0$, $\gamma$ and $K$ are considered as constant within each segment.

Under the polytropic approximation, the simplest interpolation   is transforming all the data points from EOS table into logarithmic space  to implement linear interpolation. For example, the value $(\varepsilon , p)$ between its nearest neighbors $(\varepsilon_n , p_n)$ and $(\varepsilon_{n+1} , p_{n+1})$ could be given as \cite{methodL,methodL1},
\begin{equation} \label{linearinterp}
\frac{{{{\log }_{10}}p - {{\log }_{10}}{p_n}}}{{{{\log }_{10}}\varepsilon  - {{\log }_{10}}{\varepsilon _n}}} = \frac{{{{\log }_{10}}{p_{n + 1}} - {{\log }_{10}}{p_n}}}{{{{\log }_{10}}{\varepsilon _{n + 1}} - {{\log }_{10}}{\varepsilon _n}}}.
\end{equation}
In addition to the linear interpolation, an advanced method that simultaneously preserves the monotony and the first derivative continuity of the EOS is the Piecewise-Cubic-Hermite-Interpolating-Polynomial (PCHIP) \cite{HI,PCHIP} which could be written as the follows,
\begin{eqnarray}
p\left( \varepsilon  \right) &=& {p_n}{\alpha _n}\left( \varepsilon  \right) + {p_{n + 1}}{\alpha _{n + 1}}\left( \varepsilon  \right) + {k_n}{\beta _n}\left( \varepsilon  \right) + {k_{n + 1}}{\beta _{n + 1}}\left( \varepsilon  \right),\label{pchipII}
\end{eqnarray}
where $k_{n}$ and $k_{n+1}$ are the slopes of interpolated function $p(\varepsilon)$ at $\varepsilon_n$ and $\varepsilon_{n+1}$, respectively. The specific rules to give the knot slopes $k_{n}$ are introduced in the appendix. The four Hermitian functions are given as,
\begin{eqnarray}
{\alpha _n}\left( \varepsilon  \right) &=& \left( {1 + 2\frac{{\varepsilon  - {\varepsilon _n}}}{{{\varepsilon _{n + 1}} - {\varepsilon _n}}}} \right){\left( {\frac{{\varepsilon  - {\varepsilon _{n + 1}}}}{{{\varepsilon _n} - {\varepsilon _{n + 1}}}}} \right)^2} ,\label{pchipI} \\
 \quad {\alpha _{n + 1}}\left( \varepsilon  \right) &=& \left( {1 + 2\frac{{\varepsilon  - {\varepsilon _{n + 1}}}}{{{\varepsilon _n} - {\varepsilon _{n + 1}}}}} \right){\left( {\frac{{\varepsilon  - {\varepsilon _n}}}{{{\varepsilon _n} - {\varepsilon _{n + 1}}}}} \right)^2},\\
{\beta _n}\left( \varepsilon  \right) &=& \left( {\varepsilon  - {\varepsilon _n}} \right){\left( {\frac{{\varepsilon  - {\varepsilon _{n + 1}}}}{{{\varepsilon _n} - {\varepsilon _{n + 1}}}}} \right)^2}, \\
\quad {\beta _{n + 1}}\left( \varepsilon  \right) &=& \left( {\varepsilon  - {\varepsilon _{n + 1}}} \right){\left( {\frac{{\varepsilon  - {\varepsilon _n}}}{{{\varepsilon _n} - {\varepsilon _{n + 1}}}}} \right)^2}.
\end{eqnarray}
In order to facilitate the analysis, we divide the tabular EOSs into two parts: the low density segment $[0, \rho_0]$ and the supra-nuclear segment $[\rho_0 , 10\rho_0]$. In actual operation, the minimum density  is adopted as $10^4$ $\rm{kg/m^3}$. For comparison, we define APR-a as an exact EOS example, which is spline-fitted from the well-known APR EOS \cite{APR_EoS} by using a smoothing parameter 0.989 (corresponding to a smallest R-square). To discuss the errors induced from interpolation process, we uniformly sample from APR-a at logarithmic density in the two segment to produce EOS tables in different resolution. The grid points in each density segment are logarithmic-uniform as log$\rho_{n+1}$-log$\rho_n$=constant. Up to date, most of available EOS tables are provided with density grid points of this type (denoted as $U$-grid) at high density segment \cite{stellar}. We also use a concise symbol to denote the resolution of an EOS table. For example, APR-20(200) indicates that the table of APR EOS contains totally 200 points, of which 20 points are distributed at supra-nuclear density segment.

We produce EOS tables of scale 20(200), 20(400), 50(200) and 50(100)  from APR-a EOS under $U$-grid. The relative residuals are estimated with the two interpolation methods mentioned above, and the results are present in Tab. \ref{Umesh}. It's worth pointing out
that the integration process is adequately precise, where the mass errors are at $\sim0.001\%$ level, which are negligible compared with the interpolation-induced errors.

\begin{table}[!htbp]
\centering
\begin{tabular}{|c|c|c|c|c|c|c|c|c|c|}
\hline
\multicolumn{2}{|c|}{ \multirow{2}*{Exact} } & \multicolumn{4}{c|}{APR-20(200)} & \multicolumn{4}{c|}{APR-50(200)}\\
\cline{3-10}
\multicolumn{2}{|c|}{} & \multicolumn{2}{c|}{Linear} & \multicolumn{2}{c|}{PCHIP} & \multicolumn{2}{c|}{Linear} & \multicolumn{2}{c|}{PCHIP}\\
\hline
 $\rho_c(\rho_0)$ & $M(M_{\odot})$ & $resM(\%)$ & $res\Lambda(\%)$ & $resM(\%)$ & $res\Lambda(\%)$ & $resM(\%)$ & $res\Lambda(\%)$ & $resM(\%)$ & $res\Lambda(\%)$\\
\hline
3.3377&1.0900 	&0.0669	&0.3517	&0.0887	&0.4766	&0.0186	&0.0624	&0.0135	&0.0525\\
4.0071&1.4000	&0.4114	&2.8483	&0.3859	&2.4134	&0.0458	&0.3946	&0.0284	&0.2269\\
4.6828&1.6500 	&0.2489	&1.4483	&0.1259	&0.6086	&0.0496	&0.4255	&0.0253	&0.2667\\
6.7543&2.0500 	&0.0562	&0.2848	&0.0104	&0.0797	&0.0232	&0.1820	&0.0108	&0.1404\\

\hline
\multicolumn{2}{|c|}{ \multirow{2}*{Exact} } & \multicolumn{4}{c|}{APR-20(400)} & \multicolumn{4}{c|}{APR-50(100)}\\
\cline{3-10}
\multicolumn{2}{|c|}{} & \multicolumn{2}{c|}{Linear} & \multicolumn{2}{c|}{PCHIP} & \multicolumn{2}{c|}{Linear} & \multicolumn{2}{c|}{PCHIP}\\
\hline
 $\rho_c(\rho_0)$ & $M(M_{\odot})$ & $resM(\%)$ & $res\Lambda(\%)$ & $resM(\%)$ & $res\Lambda(\%)$ & $resM(\%)$ & $res\Lambda(\%)$ & $resM(\%)$ & $res\Lambda(\%)$\\
\hline
3.3377&1.0900 	&0.0672	&0.3788	&0.0894	&0.4812	&0.0298	&0.8041	&0.0130	&0.3395\\
4.0071&1.4000	&0.4066	&2.8416	&0.3816	&2.4207	&0.0465	&0.4866	&0.0404	&0.3764\\
4.6828&1.6500 	&0.2313	&1.4434	&0.1251	&0.6026	&0.0537	&0.5554	&0.0408	&0.4020\\
6.7543&2.0500 	&0.0562	&0.2882	&0.0103	&0.0758	&0.0208	&0.2742	&0.0114	&0.0917\\

\hline
\end{tabular}
\caption{Relative residuals for the two interpolation methods. The most left column ($\rho_c$) is the central density for each corresponding row. The relative residuals $resM$ and $res\Lambda$ are defined as the relative deviations of $M$ and $\Lambda$ to their exact solutions, respectively.}\label{Umesh}
\end{table}

As we know, the mass results are most sensitive to the EOS of the star's central region. As shown in Tab.$\,$\ref{Umesh}, the relative smaller $resM$ of PCHIP indicates that the PCHIP interpolation have advantages over the linear interpolation.

The precision of $\Lambda$ is affected simultaneously by $k_2$, $M$ and $R$. Under the $U$-grid the precision of $k_2$ is generally on the same order of $M$ and their relative residuals can be estimated as rough views of the hydrostatic solution precision. The   error of $R$ is mainly related to the integration step-size which we will not discuss in detail here, but as a conclusion the relative error of $R$ is $\sim 0.01\%$ for the four EOSs in Tab. \ref{Umesh}. Thus it is easy to understand that as the $resR \ll resM$ and $resk_2 \sim resM$ for APR-20(200) and APR-20(400), the relative residual of Love-number is $\sim 6 \, resM$.

Our calculation shows that for both the PCHIP and linear interpolation, the EOS tables with denser data-points at high density side would significantly decrease both the mass residuals and the Love number residuals. The data-point amount of APR-20(200) is the same as APR-50(200), but the APR-50(200) produces much more precise outcomes because of the better resolution at high density side. On the other hand, through comparing the data of APR-20(400) and APR-20(200) in Tab.$\,$\ref{Umesh}, it is shown that the resolution improvement at low density  can not remarkably reduce the interpolation-induced errors.

The comparisons indicate that the resolution of EOS table at supra-nuclear density is much more important than the total amount of data-points. In addition, we notice that  too small-size EOS table, such as APR-50(100), would violate the approximation that $resk_2 \sim resM$, and thus reduce the $\Lambda$ precision, especially for the low-mass stars. Through more trials on the resolution at both density segments, the minimal scale for $U$-grid APR EOS is finally determined as 50(150), which corresponds to the interpolation-induced errors $\sim 0.02\%$ for $M$ and $\sim0.2\%$ for $\Lambda$.

Actually, lots of EOSs in the literatures prefer to adopt the table scale around 20(200) under $U$-grid \cite{stellar}, such as SFHo, GShen and LS EOSs \cite{SFHo,GShen,LS}. The interpolation-induced errors for the stellar mass $M$ produced by these EOS tables are expected to be $0.1\% \sim 1\%$. According to above discussion, it is suggested that when we produce the EOS table, it is better to contain more than 50 grid points at supra-nuclear density segment [$\rho_0, 10\rho_0$]. It would significantly reduce the interpolation-induced errors and thus result in much more precise solutions for such as the stellar mass $M$ and tidal deformability $\Lambda$.

\section{Dependence on grid specification}

To eliminate accidental factors, we further inspect the interpolation errors under several different grid modes, such as
\begin{alignat}{2} \label{Def_grid}
& \frac{{{{\log }_{10}}{\rho _{n + 1}} - {{\log }_{10}}{\rho _n}}}{{{{\log }_{10}}{\rho _n} - {{\log }_{10}}{\rho _{n - 1}}}}= \begin{cases}  {C_1} , \\   {\frac{{1 + {e^{ - {C_2} \cdot (n + 1)}}}}{{1 + {e^{ - {C_2} \cdot n}}}}} , \end{cases}
& &
\begin{aligned}
\text{for } {G{\rm{ - grid}}} \\
\text{for } {ue{\rm{ - grid}}}
\end{aligned}
\end{alignat}
where $C_1$ and $C_2$ are adjustable coefficients to meet the resolution requirement. Apparently, taking $C_1=1$ for $G$-grid is equivalent to $U$-grid we mentioned in Sec.$\,$II.

\begin{figure}
\centering
\includegraphics [width=0.6\textwidth]{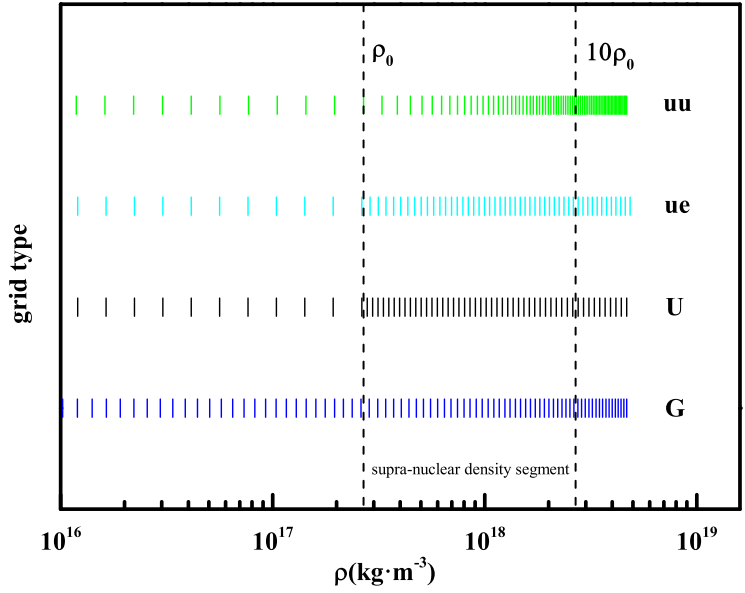}
\caption{ \label{GridType} Baryon density grid points for APR-50(150) under four different grid modes.}
\end{figure}

Four different grid modes are defined as the follows. (i) To $uu$-grid, we separately take $C_1=1$ at the low density segment and $C_1=0.1$ at the high density segment. (ii) To $ue$-grid, we adopt $C_2=0.56$ at the supra-nuclear density segment but share the same grids from $uu$-grid at the low density segment. (iii) To $U$-grid, the EOS table is logarithmic-uniform ($C_1=1$) to consistent with Sec. II. (iv)  To $G$-grid, the constant should be $C_1=0.9785$ at  both density segments. A concise example of the four grids for APR-50(150) is plotted in Fig. \ref{GridType}.

The $uu$-grid means uniform distribution at [$\rho_0, 10\rho_0$] and logarithmic-uniform at [$0, \rho_0$]. It is an extreme distribution that the grid points are concentrated excessively at the high density side. The $U$-grid could be considered as the opposite extreme with respect to the $uu$-grid. Any distribution that less-contractive than $U$-grid should be irrational according to the discussion in Sec. II. From the definition of $G$-grid and $ue$-grid, by contrast, they are almost transitional schemes between $U$-grid and $uu$-grid. The $ue$-grid, different from the frameworks of $G$-grid, is designed to restrain the contractive rate that the final interval of the logarithmic grid is expected to be about half of the beginning one.

\begin{figure}
\centering
\includegraphics [width=0.6\textwidth]{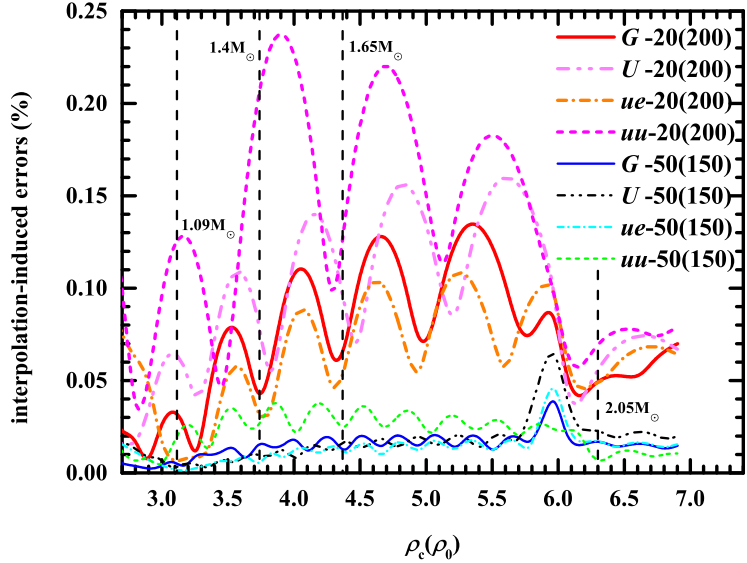}
\caption{ \label{EOSdispMethod} Residuals of  APR-20(200) and APR-50(150) under four different mesh grids and with the PCHIP interpolation. The interpolation-induced errors are defined as the relative deviations of $M$ to the exact solutions.}
\end{figure}

According to the residuals comparison of the four meshing methods in Fig.$\,$\ref{EOSdispMethod}, it is clear that the interpolation-induced errors are related to both the choice of the EOS mesh and the central density. For a neutron star, if its central density is near the data-point in the EOS tables, it would naturally correspond to a smaller error. For example, the $uu$-grid that most stressed on the high density side results in the most precise outcomes at $\sim 2 \, \rm{M}_{\odot}$ but the least precise outcomes at low-mass side. Additionally, the interpolation errors of EOS itself are generally small nearby the data-point and achieving maximum at the intermediate position before the next point. These regular changes lead to oscillatory contours in Fig.$\,$\ref{EOSdispMethod}.

Considering both the practicability and the overall accuracy, the widely-used $U$-grid remains the optimal choice to produce EOS tables. Although there is certain meshing-method dependence in accuracy issue, we may still conclude from Tab.$\,$\ref{Umesh} and Fig.$\,$\ref{EOSdispMethod} that the EOS tables of scale 50(150) could effectively restrain the interpolation-induced errors of stellar mass to $resM\sim 0.02\%$ within $1.09 \, \rm{M}_{\odot}<M<2.05 \, \rm{M}_{\odot}$.

Someone may doubt the universality of the conclusion because all the EOS tables in comparison are sampled from a single EOS model. To make the conclusion more reliable, we extend the same interpolation trials on the parameterized asymmetric nucleon matter EOSs \cite{LiEoS1}. We generated several tens of EOS tables in the resolution of 50(150), and then separately estimated the interpolation errors with different methods. We found that it is consistent with the former analyses of the APR EOS.

\section{summary and conclusion}

We reinvestigated the traditional numeric methods for computing TOV equation including both the integration techniques and the interpolation methods.
For the convenience of discussion, we separated the global errors into integration errors and interpolation-induced errors. As the integration residuals near the stellar surface are divergent and the radial precision can be affected strongly by the choice of step-size, the adaptive step-size method is adopted to solve this problem.

Currently, a bulk of available EOS tables are provided in the general scale of $\sim20(200)$ \cite{APR_EoS,stellar,SFHo,GShen,LS}. As the errors from integration process can be well-handled, the dominant errors of using these EOS tables would come from the interpolation process. The relation between the interpolation-induced errors and the EOS table resolution under $U$-grid are investigated in detail. It is concluded that the increasing amount of data-points at supra-nuclear density segment [$\rho_0, 10 \rho_0$] could effectively reduce the interpolation-induced errors. The EOS table of scale 50(150) would correspond to the relative residuals for $M$ and $\Lambda$ to $\sim0.02\%$ and $\sim0.2\%$, respectively, which is much more accurate than the 20(200) ones. In addition, it is also shown  that the PCHIP method is more accurate than the linear method.

The dependence of the meshing methods is finally inspected. Among the four specified meshing methods, $U$-grid remains the optimal method for generating EOS tables to compute intermediate-mass neutron stars. The EOS-model dependence is also examined by  the parameterized asymmetric nucleonic matter EOSs \cite{LiEoS1}. It is concluded that EOS tables of scale 50(150) could still significantly improve the accuracy compared with the general scale 20(200), despite of certain differences in meshing methods, EOS models or interpolation methods. By the way, all of the source programs (C codes) are already publicly available, please refer to Ref. \cite{C_codes}.

\begin{acknowledgements}
We would like to thank Bao-An Li for helpful discussions. This work is supported by the National Natural Science Foundation of China (No.11722546 and No.11275073), talent program of South China University of Technology (No. K5180470). This project is sponsored by CSC and has made use of NASA's Astrophysics Data System.

\end{acknowledgements}


\appendix

\section{piecewise-cubic-hermite-interpolation}
Varying with the specific method to give the boundary slopes $k_n$ and $k_{n+1}$ at Eqs. (\ref{pchipI}-\ref{pchipII}), there are countless executions of Hermite interpolation, and the one used in this paper is a specified weighted average of differential slopes as its nodal slopes $k_n$ to preserve the shape of interpolated function from the origin data-points. The expression of nodal slopes $k_n$ for inner-points and endpoints are separately given in Eqs. (\ref{pchipinnerpoint})-(\ref{pchipendpoint}).

Generally, $k_{n}$ at each knot is uniquely determined by differential aspects of proximal points. Hermite interpolation therefore ensures the first derivative $dp/d\varepsilon$ is continuous everywhere. We denote the right differential step of $\varepsilon_n$ as $h_n=\varepsilon_{n+1} - \varepsilon_n$ and the right differential slope as $\nu_n=(p_{n+1}-p_n)/h_n$. When $sgn(\nu_{n-1}) \ne sgn(\nu_{n})$, we choose $k_n = 0$ so that the extremum of data-points could be coincident with that of interpolated function, although it is not very likely to have $sgn(\nu_{n-1}) \ne sgn(\nu_{n})$ within a rigorous EOS. In the general case, $k_n$ of the inner-points are given as,
\begin{equation} \label{pchipinnerpoint}
\begin{array}{l}
\widetilde {{\nu _n}} = \frac{{{h_{n - 1}}{\nu _{n - 1}} + {h_n}{\nu _n}}}{{{h_{n - 1}} + {h_n}}}\\
{k_n} = \frac{{3{\nu _{n - 1}}{\nu _n}}}{{{\nu _{n - 1}} + {\nu _n} + \widetilde {{\nu _n}}}}
\end{array}
\end{equation}

Slopes estimation at the two endpoints $k_1$ and $k_{end}$ are slightly different from that at inner-points. We denote the differential step from an endpoint to its nearest neighbour as $h_{1*}=\varepsilon_{1*}-\varepsilon_{1,end}$, and denote that from nearest neighbour to sub neighbour as $h_{2*}=\varepsilon_{2*}-\varepsilon_{1*}$, while the corresponding differential slopes are $\nu _{1*}=(p_{1*}-p_{1,end})/h_{1*}$ and $\nu _{2*}=(p_{2*}-p_{1*})/h_{2*}$. In the estimation, we firstly give $k_{1,end}$ as Eq.$\,$(\ref{pchipendpoint}). If $sgn(k_{1,end}) \ne sgn(\nu_{1*})$ we choose $k_{1,end}=0$, else if $sgn(\nu_{1*}) \ne sgn(\nu_{2*})$ and $\left| k_{1,end} \right| > \left| 3\nu_{1*} \right|$ we choose $k_{1,end}=3\nu_{1*}$, and only when all the judgements above are false $k_{1,end}$ remains invariant.
\begin{equation} \label{pchipendpoint}
{k_{1,end}} = \frac{{\left( {2{h_{1*}} + {h_{2*}}} \right){\nu _{1*}} - {h_{1*}}{\nu _{2*}}}}{{{h_{1*}} + {h_{2*}}}}
\end{equation}

The specific rules to give $k_{n}$ above are designed to preserve monotony and avoid overshooting, as the interpolated EOSs are generally expected to be not oscillatory and baratropic.

\end{document}